\documentclass[sigconf]{acmart}

\usepackage[overload]{textcase} 
\usepackage{multirow}
\usepackage{color}

\usepackage{amsmath,amssymb,bm}
\usepackage{graphicx}
\usepackage{float}
\usepackage{listings}

\copyrightyear{2026}
\acmYear{2026}
\setcopyright{cc}
\setcctype{by}
\acmConference[SIGIR '26]{Proceedings of the 49th International ACM SIGIR Conference on Research and Development in Information Retrieval}{July 20--24, 2026}{Melbourne, VIC, Australia}
\acmBooktitle{Proceedings of the 49th International ACM SIGIR Conference on Research and Development in Information Retrieval (SIGIR '26), July 20--24, 2026, Melbourne, VIC, Australia}
\acmDOI{10.1145/3805712.3808560}
\acmISBN{979-8-4007-2599-9/2026/07}

\begin{document}

\title{A Reproducibility Study of LLM-Based Query Reformulation}

\author{Amin Bigdeli}
\orcid{0009-0003-8977-9312}
\affiliation{%
  \institution{University of Waterloo}
  \city{Waterloo}
  \state{Ontario}
  \country{Canada}
}
\email{abigdeli@uwaterloo.ca}

\author{Radin Hamidi Rad}
\orcid{0000-0002-9044-3723}
\affiliation{%
     \institution{Mila – Quebec AI Institute}
     \city{Montreal}
     \state{Quebec}
     \country{Canada}
}
\email{radin.hamidi-rad@mila.quebec}

\author{Hai Son Le}
\orcid{0009-0003-2240-0451}
\affiliation{%
  \institution{Toronto Metropolitan University}
  \city{Toronto}
  \state{Ontario}
  \country{Canada}
}
\email{haison.le@torontomu.ca}

\author{Mert Incesu}
\orcid{0009-0007-1129-2597}
\affiliation{%
  \institution{University of Toronto}
  \city{Toronto}
  \state{Ontario}
  \country{Canada}
}
\email{mert.incesu@mail.utoronto.ca}

\author{Negar Arabzadeh}
\orcid{0000-0002-4411-7089}
\affiliation{%
  \institution{University of California, Berkeley}
  \city{Berkeley}
  \state{California}
  \country{USA}
}
\email{negara@berkeley.edu}

\author{Charles L. A. Clarke}
\orcid{0000-0001-8178-9194}
\affiliation{%
  \institution{University of Waterloo}
  \city{Waterloo}
  \state{Ontario}
  \country{Canada}
}
\email{charles.clarke@uwaterloo.ca}

\author{Ebrahim Bagheri}
\orcid{0000-0002-5148-6237}
\affiliation{%
  \institution{University of Toronto}
  \city{Toronto}
  \state{Ontario}
  \country{Canada}
}
\email{ebrahim.bagheri@utoronto.ca}

\renewcommand{\shortauthors}{Amin Bigdeli et al.}

\begin{abstract}
Large Language Models (LLMs) are now widely used for query reformulation and expansion in Information Retrieval, with many studies reporting substantial effectiveness gains. However, these results are typically obtained under heterogeneous experimental conditions, making it difficult to assess which findings are reproducible and which depend on specific implementation choices. In this work, we present a systematic reproducibility and comparative study of ten representative LLM-based query reformulation methods under a unified and strictly controlled experimental framework. We evaluate methods across two architectural LLM families at two parameter scales, three retrieval paradigms (lexical, learned sparse, and dense), and nine benchmark datasets spanning TREC Deep Learning and BEIR. Our results show that reformulation gains are strongly conditioned on the retrieval paradigm, that improvements observed under lexical retrieval do not consistently transfer to neural retrievers, and that larger LLMs do not uniformly yield better downstream performance. These findings clarify the stability and limits of reported gains in prior work. To enable transparent replication and ongoing comparison, we release all prompts, configurations, evaluation scripts, and run files through QueryGym, an open-source reformulation toolkit with a public leaderboard.\footnote{\url{https://leaderboard.querygym.com}}

\end{abstract}




\maketitle

\section{Introduction}
\label{sec:intro}

Query reformulation and expansion have historically served as key mechanisms for bridging the gap between how users express their information needs and how retrieval systems represent documents \cite{mo2023convgqr,arabzadeh2021matches,deutsch2006query,bigdeli2024learning}. Traditional approaches, including pseudo-relevance feedback~\cite{rocchio1971relevance}, ontology-based expansion~\cite{bhogal2007review}, and concept-driven enrichment~\cite{qiu1993concept}, operate by injecting additional terms into the query from external knowledge sources or top-ranked documents \cite{lu2015query,pal2014improving,dang2010query}. While these methods have proven effective under certain conditions, they remain vulnerable to noisy feedback and topic drift, often producing inconsistent improvements across queries and collections~\cite{abdul2004umass,boldi2011query,ooi2015survey,hosseini2024enhanced}.

The emergence of LLMs has introduced a fundamentally different paradigm for query reformulation, one in which a generative model draws on its parametric knowledge to rewrite or augment a query without requiring access to initial retrieval results or structured knowledge bases. This shift has given rise to a diverse and rapidly growing family of methods~\cite{query2doc,mugi,lamer,bigdeli2026reformer,Genqr,Genqrensemble,csqe,qa-expand}. Instruction-driven approaches such as \texttt{GenQR}~\cite{Genqr} and \texttt{GenQREnsemble}~\cite{Genqrensemble} prompt an LLM to generate high-impact expansion terms directly from the query. Pseudo-document strategies, exemplified by \texttt{Query2Doc}~\cite{query2doc} and \texttt{MUGI}~\cite{mugi}, synthesize answer-like passages that are appended to the original query to enrich its lexical and semantic coverage. Question-answer pipelines such as \texttt{QA-Expand}~\cite{qa-expand} decompose a query into sub-questions and selectively fold generated answers back into the expanded representation. Corpus-grounded methods including \texttt{LameR}~\cite{lamer} and \texttt{CSQE}~\cite{csqe} condition generation on initially retrieved evidence or collection-level statistics to anchor the expansion in distributional properties of the target corpus. A broader survey of these families and their design trade-offs is provided by~\citet{mugi} and~\citet{exploringresearch}.

Despite many proposed approaches, evaluation settings vary widely across studies, creating a fragmented landscape that hinders reproducibility and fair comparison. Individual papers typically select a single LLM backbone, apply method-specific prompting templates, adopt particular decoding configurations including temperature, sampling strategy, and maximum token length, and evaluate on a narrow set of benchmarks using a single retrieval method. Because these choices are rarely held constant across studies, it is difficult to determine whether reported improvements stem from the reformulation strategy, the LLM used, the decoding parameters, or the retrieval infrastructure. Prior analyses of evaluation practices in neural IR have highlighted the risks of such uncontrolled comparisons~\cite{lin2019neural,yang2019critically}, and the proliferation of LLM-based methods has only amplified these concerns where the number of confounding variables has grown, while standardized evaluation protocols have not kept pace.

Beyond this lack of experimental standardization, two substantive gaps limit the current understanding of LLM-based query reformulation. \textbf{First}, the overwhelming majority of existing evaluations pair reformulation methods exclusively with lexical retrievers. While this provides a well-understood baseline, it leaves open the question of whether reformulation gains transfer to retrieval paradigms that operate on fundamentally different representational assumptions. Learned sparse models already perform neural term expansion during encoding, potentially rendering additional query-side expansion redundant or even counterproductive. Dense retrievers encode queries and documents into continuous vector spaces where surface-level lexical modifications may not translate into proportional shifts in embedding similarity. The interaction between the representation space of the retriever and the modifications introduced by generative reformulation is therefore a critical factor that existing work leaves largely uncontrolled. \textbf{Second}, the recent trend toward increasingly large LLMs introduces a cost-effectiveness trade-off that has not been systematically characterized for query reformulation. A large-scale model incurs substantially greater computational cost and latency than a compact alternative within the same architectural family, yet whether the resulting reformulations yield proportionally better retrieval outcomes remains an open question. Answering this question requires controlled within-family comparisons at different parameter scales under identical experimental conditions, rather than ad-hoc inferences drawn across studies that differ along multiple axes simultaneously. To enable this controlled comparison, all reformulation methods are executed through QueryGym~\cite{querygym}, a unified open-source toolkit that provides standardized implementations under a shared prompting and decoding interface.

In this paper, we address these gaps through a systematic reproducibility and comparative study that jointly controls the factors most commonly conflated in existing work. We evaluate ten representative reformulation methods, spanning instruction-based, pseudo-document, question-answer, and corpus-grounded families, using LLMs from two architectural lineages (GPT-4.1~\cite{gpt4.1} and Qwen2.5~\cite{qwen25}), each at two parameter scales (small and large). All models are executed under identical decoding configurations, including maximum token length and sampling parameters, ensuring that observed differences in retrieval effectiveness can be attributed to the reformulation method and the model characteristics rather than to hidden generation-level variance. We assess downstream retrieval impact across three paradigms, namely lexical, learned sparse, and dense retrieval. To evaluate generalization and domain sensitivity, our experiments span nine benchmarks including the TREC Deep Learning collections~\cite{TREC2019,TREC2020,DL_HARD}, which represent large-scale web search with graded relevance judgments, and six datasets from the BEIR benchmark~\cite{kamalloo}, covering scientific, argumentative, biomedical, financial, encyclopedic, and news domains. Throughout the study, preprocessing pipelines, prompting protocols, indexing configurations, and evaluation metrics are held constant to isolate the contribution of reformulation strategies from confounding implementation effects.

This reproducibility study is organized around four research questions, each targeting a dimension that is frequently conflated in existing literature:

\begin{itemize}
    \item \textbf{RQ1 (Cross-Method Comparison):} Under identical experimental conditions, how do different LLM-based reformulation methods compare in terms of retrieval effectiveness, robustness across settings, and sensitivity to model and retriever choice?
    \item \textbf{RQ2 (Cross-Retriever Performance):} How do reformulation gains interact with the underlying retrieval paradigm, and do methods that improve lexical retrieval also benefit learned sparse and dense retrievers?
    \item \textbf{RQ3 (Domain Robustness and Dataset Sensitivity):} To what extent do LLM-based query reformulation methods generalize across datasets with heterogeneous topical domains and query characteristics, and which methods exhibit the greatest sensitivity to domain shift?
    \item \textbf{RQ4 (LLM Backbone and Scale):} What is the impact of the LLM's architectural lineage and parameter scale on the quality of generated reformulations and their downstream retrieval effectiveness?
\end{itemize}

These research questions collectively enable a multi-dimensional evaluation that disentangles the contribution of the reformulation strategy from the choice of LLM, retrieval architecture, and evaluation domain. Rather than proposing a new reformulation technique, our objective is to establish which reported findings hold under controlled conditions and to identify where the boundaries of current methods lie. The main contributions of this work are threefold:

\begin{itemize}
    \item \textbf{Controlled cross-method comparison.} We present a systematic reproducibility study of LLM-based query reformulation methods that enforces unified decoding configurations across all models and methods, eliminating hidden variance stemming from inconsistent temperature, token limits, and sampling settings, and enabling the first fair head-to-head comparison of representative reformulation families under identical conditions.
    \item \textbf{Cross-paradigm retriever analysis.} We extend the evaluation of LLM-driven reformulation beyond lexical retrieval to learned sparse and dense paradigms, providing empirical evidence on how reformulation gains vary, and in some cases reverse, across fundamentally different representation spaces.
    \item \textbf{Multi-domain benchmarking with reproducible artifacts.} We deliver a large-scale comparative evaluation across nine benchmark datasets from the TREC Deep Learning and BEIR collections, characterizing domain sensitivity and accompanied by all prompts, configurations, evaluation scripts, and a public leaderboard released through QueryGym to support transparent replication and future extension.
\end{itemize}

Our findings indicate that the effectiveness of LLM-based reformulation is more conditional than aggregate metrics in prior studies suggest. While several methods produce reliable gains under lexical retrieval, these benefits frequently diminish or reverse when the same reformulations are issued to learned sparse or dense retrievers. The relationship between model scale and downstream effectiveness is similarly nuanced where larger LLMs do not uniformly outperform their compact counterparts, and the magnitude and direction of scale effects depend on both the reformulation method and the target domain. These observations suggest that a number of improvements reported in prior work are tied to specific evaluation configurations and do not generalize across the retriever paradigms and domain conditions examined in this study.

\section{Experimental Setup}
\label{sec:setup}

\subsection{Datasets}

Our evaluation spans nine benchmark datasets selected to cover both high-resource web search and domain-diverse retrieval conditions.

\noindent \textbf{TREC Deep Learning.} We use the TREC Deep Learning test collections: TREC DL 2019~\cite{TREC2019}, TREC DL 2020~\cite{TREC2020}, and DL-HARD~\cite{DL_HARD}. All three query sets are derived from the MS~MARCO V1 passage collection~\cite{nguyen2016ms} and are accompanied by graded relevance judgments over large candidate pools. DL-HARD specifically targets queries that are particularly challenging for standard retrieval systems, enabling evaluation under both typical and adversarial web search conditions.

\noindent \textbf{BEIR.} To assess cross-domain robustness, we incorporate six datasets from the BEIR benchmark~\cite{kamalloo}: SciFact, ArguAna, COVID, FiQA, DBPedia, and News, spanning scientific, argumentative, biomedical, financial, encyclopedic, and news domains. Their inclusion enables evaluation of whether reformulation gains observed on the MS~MARCO distribution transfer to heterogeneous retrieval settings.

\noindent Table~\ref{tab:datasets} reports the number of queries, corpus size, and domain for each dataset.

\begin{table}[t]
\centering
\caption{Overview of the evaluation datasets.}
\vspace{-1em}
\label{tab:datasets}
\small
\begin{tabular}{llrrr}
\toprule
\textbf{Source} & \textbf{Dataset} & \textbf{\#Queries} & \textbf{\#Documents} & \textbf{Domain} \\
\midrule
\multirow{3}{*}{TREC DL} & DL 2019 & 43 & \multirow{3}{*}{8,841,823} & \multirow{3}{*}{Web Search} \\
 & DL 2020 & 54 &  &  \\
 & DL-HARD & 50 &  &  \\
\midrule
\multirow{6}{*}{BEIR} & SciFact & 300 & 5,183 & Scientific \\
 & ArguAna & 1,406 & 8,674 & Argumentative \\
 & COVID & 50 & 171,332 & Bio-Medical  \\
 & FiQA & 648 & 57,638 & Financial \\
 & DBPedia & 400 & 4,635,922 & Wikipedia \\
 & News & 57 & 594,977 & News \\
\bottomrule
\end{tabular}
\vspace{-1em}
\end{table}

\subsection{Query Reformulation Methods}

We evaluate various LLM-based query reformulation methods that represent the major methodological families in current literature. For each original query $q$, a method produces an expanded or reformulated query $q'$, which is then issued to the retrieval pipeline. All methods are executed using the same LLM under identical decoding configurations, and no method-specific tuning or retriever-specific adjustments are applied across experimental conditions. This protocol ensures that observed performance differences reflect the reformulation strategy rather than implementation-level confounds. We organize the evaluated methods into three categories based on their underlying reformulation strategy.

\subsubsection{Keyword-Level Expansion}
Methods in this category prompt the LLM to generate additional keywords or phrases that are appended to the original query, operating at the term level without synthesizing extended passages or conditioning on external signals.

\begin{itemize}

    \item \textbf{\texttt{GenQR}}~\cite{Genqr}: Prompts the LLM with an instruction to produce high-impact expansion terms that are appended to the original query. It operates in a zero-shot setting with no retrieval feedback or in-context examples and generates keywords through $N{=}5$ independent LLM calls.

    \item \textbf{\texttt{GenQREnsemble}}~\cite{Genqrensemble}: Extends \texttt{GenQR} by paraphrasing the reformulation instruction into ten lexically diverse variants and issuing each independently to the LLM. The resulting keyword sets from all 10 instructions are merged into a single consolidated query, exploiting prompt diversity to elicit complementary expansion terms.

    \item \textbf{\texttt{Query2Keyword (Q2K)}}~\cite{exploringresearch}: Maps the original query to an explicit expanded representation by prompting the LLM in a single call to generate semantically related terms and phrases, aiming to broaden lexical coverage without relying on pseudo-document synthesis or question decomposition.

\end{itemize}

\subsubsection{Document-Level Expansion}
Methods in this category prompt the LLM to synthesize extended textual content, such as answer-style passages or responses to decomposed sub-questions, which is then integrated with the original query to provide richer semantic and topical signals.

\begin{itemize}

    \item \textbf{\texttt{Query2Doc (Q2D)}}~\cite{query2doc}: Synthesizes an answer-style pseudo-document conditioned on the original query in a single LLM call and concatenates it with the query to enrich lexical and semantic coverage. We evaluate three prompting variants: \emph{zero-shot} (\texttt{ZS}), \emph{few-shot} (\texttt{FS}) with in-context examples, and \emph{chain-of-thought} (\texttt{CoT}) with intermediate reasoning.

    \item \textbf{\texttt{QA-Expand}}~\cite{qa-expand}: Generates three diverse sub-questions from the original query, produces corresponding pseudo-answers for each, and applies a feedback-driven rewriting and filtering stage that retains only the most informative answers. The refined pseudo-answers are then concatenated with the original query as expansion content.

    \item \textbf{\texttt{MUGI}}~\cite{mugi}: Generates five independent pseudo-documents from the original query and consolidates them into a single expanded representation, increasing diversity and coverage while mitigating noise from any single generation. The method adaptively weights the original query against the generated content to balance lexical emphasis.

\end{itemize}

\subsubsection{Corpus-Grounded Expansion}
Methods in this category condition the generation process on signals from the target collection or initial retrieval results, anchoring the expansion in corpus-level distributional properties rather than relying solely on the model's parametric knowledge.

\begin{itemize}

    \item \textbf{\texttt{CSQE}}~\cite{csqe}: Generates two LLM passages steered by collection-level distributional statistics and supplements them with relevant sentences extracted from the top-10 retrieved documents via LLM-based relevance judgments. This grounding mechanism aligns the expansion with the vocabulary and topical distribution of the target corpus.

    \item \textbf{\texttt{LameR}}~\cite{lamer}: Retrieves the top-10 documents for the original query and conditions the LLM on this evidence to produce five independent rewrites enriched with disambiguating context and descriptive language.

\end{itemize}

\subsection{Large Language Models}

To systematically examine the impact of architectural lineage and parameter scale on reformulation quality, we select LLMs from two distinct model families, each represented at two parameter scales. From the proprietary GPT family~\cite{gpt4.1}, we employ GPT-4.1 as the large-scale variant and GPT-4.1-nano as the compact variant. From the open-weight Qwen family~\cite{qwen25}, we employ Qwen2.5-72B as the large-scale variant and Qwen2.5-7B as the compact variant. This $2 \times 2$ design (two families $\times$ two scales) enables two types of controlled comparison: \textit{within-family} comparisons that isolate the effect of parameter scale while holding architectural lineage constant, and \textit{across-family} comparisons at matched scale that reveal the influence of model design and training methodology. All decoding configurations are detailed in Section~\ref{sec:implementation}.

\subsection{Retrieval Methods}

To evaluate whether reformulation gains are consistent across retrieval architectures or are dependent on the representation space of the retriever, we employ three different first-stage retrieval paradigms.

\noindent \textbf{Lexical Retrieval.} We use BM25~\cite{ma2022document} as the representative lexical baseline, which scores documents based on exact term overlap weighted by inverse document frequency and document length normalization. BM25 is the most widely adopted first-stage retriever in LLM-based reformulation studies and serves as the primary reference point for cross-method comparison.

\noindent \textbf{Learned Sparse Retrieval.} We employ SPLADE~\cite{formal2022distillation} as a learned sparse retriever that maps queries and documents into high-dimensional sparse representations via neural encoders. Because SPLADE already performs implicit term expansion during encoding, it provides a particularly informative setting for assessing whether explicit LLM-based query expansion yields additional gains or introduces redundancy.

\noindent \textbf{Dense Retrieval.} We use BGE~\cite{xiao2024c} as the dense retrieval model, which encodes queries and documents into continuous vector embeddings and ranks candidates based on embedding similarity. Dense retrieval operates in a representation space where surface-level lexical changes may not produce proportional shifts in similarity, making it a critical test of whether generative reformulations transfer beyond term-matching paradigms.

All three retrievers are applied with fixed indexing and ranking configurations throughout the study. No retriever-specific parameter tuning is performed, ensuring that the analysis isolates the effect of the reformulation strategy from infrastructure-level variance.

\subsection{Evaluation Metrics}

We adopt standard effectiveness metrics consistent with established reporting conventions for each benchmark family. For the TREC Deep Learning collections, we report nDCG@10 and Recall@1000 following official evaluation practices~\cite{TREC2019,TREC2020,gao2021complement}. For the BEIR datasets, we report nDCG@10 and Recall@100 in line with commonly adopted cross-domain evaluation protocols~\cite{kamalloo,zhang2023miracl}.

\subsection{Implementation Details}
\label{sec:implementation}

All reformulation methods are executed through \texttt{QueryGym}~\cite{querygym},\footnote{\url{https://github.com/ls3-lab/QueryGym}} a unified, open-source toolkit that integrates a wide range of LLM-based query reformulation methods under a shared prompting and decoding interface. \texttt{QueryGym} interfaces with any OpenAI-compatible LLM client, allowing all reformulation methods to be executed under identical generation configurations, including temperature, maximum output tokens, and prompting protocol. By unifying the generation backend across methods, \texttt{QueryGym} eliminates implementation-level and decoding-level variance as sources of confounding effects, ensuring that observed differences in retrieval effectiveness can be attributed to the reformulation method itself. The toolkit further decouples the reformulation stage from the retrieval backend, enabling the same expanded queries to be issued to lexical, learned sparse, and dense retrievers without modification. Building on \texttt{QueryGym} thus allows our study to leverage standardized implementations of every reformulation method considered, under strictly identical generation conditions across LLM backbones and retrievers. To complement the reformulation pipeline, we also release the complete set of retrieval experiment scripts, evaluation protocols, and run files used in this study, enabling end-to-end reproduction of all reported results across the three retrieval paradigms and nine benchmark datasets.\footnote{\url{https://github.com/ls3-lab/QueryGym/tree/main/reproducibility/scripts}}

For the GPT-4.1 family, we access models through the OpenAI API\footnote{\url{https://platform.openai.com/}}. For the Qwen2.5 family, we access both model scales through OpenRouter\footnote{\url{https://openrouter.ai/}}. Across all models and all reformulation methods, we set the temperature to the value recommended by each method and the maximum output token length to $256$. These settings are held strictly constant throughout the study to ensure that any observed differences in downstream retrieval effectiveness can be attributed to the reformulation method and model characteristics rather than to hidden generation-level variance.

For retrieval, all experiments are conducted using the Pyserini toolkit~\cite{pyserini}, which provides reproducible implementations of BM25, SPLADE, and BGE retrievers. For each reformulation method, the expanded query $q'$ is issued to all three retrieval pipelines over the same pre-built indexes, ensuring that retrieval-side configurations remain identical across all experimental conditions.

\section{Findings}

In this section, we present the experimental results organized around the four research questions defined in Section~\ref{sec:intro}, progressing from cross-method comparison to cross-retriever analysis, domain robustness, and the impact of LLM backbone and scale.
\subsection{RQ1. Comparative Performance of Reformulation Methods}

\begin{table}[t]
\centering
\caption{Reproducibility results for LLM-based query reformulation methods using GPT-4.1 with BM25 retrieval on TREC DL benchmarks. We report nDCG@10 and Recall@1000. Bold indicates the best score per metric. Document-generation approaches (\texttt{Q2D} variants, \texttt{MUGI}) substantially outperform  keyword-level methods, while corpus-grounded methods (\texttt{CSQE}, \texttt{LameR}) are consistently competitive across all datasets.}
\label{tab:bm25-trec}
\resizebox{\columnwidth}{!}{%
\begin{tabular}{l@{\hskip 4pt}c@{\hskip 4pt}c@{\hskip 6pt}c@{\hskip 4pt}c@{\hskip 6pt}c@{\hskip 4pt}c}
\toprule
& \multicolumn{6}{c}{\textbf{TREC DL Benchmark}} \\
\cmidrule(lr){2-7}
 & \multicolumn{2}{c}{DL 2019} & \multicolumn{2}{c}{DL 2020} & \multicolumn{2}{c}{DL-HARD} \\
\cmidrule(lr){2-3} \cmidrule(lr){4-5} \cmidrule(lr){6-7}
\textbf{Method} & nDCG@10 & R@1k & nDCG@10 & R@1k & nDCG@10 & R@1k \\
\midrule
\texttt{Original Query} & 0.506 & 0.750 & 0.480 & 0.786 & 0.285 & 0.681 \\
\midrule
\quad + \texttt{RM3} & 0.522 & 0.814 & 0.490 & 0.824 & 0.251 & 0.716 \\
\quad + \texttt{GenQR} & 0.548 & 0.828 & 0.537 & 0.840 & 0.292 & 0.743 \\
\quad + \texttt{GenQREnsemble} & 0.559 & 0.869 & 0.553 & 0.861 & 0.270 & 0.777 \\
\quad + \texttt{Q2K} & 0.594 & 0.870 & 0.576 & 0.859 & 0.345 & 0.764 \\
\midrule
\quad + \texttt{QA-Expand} & 0.683 & 0.850 & 0.642 & 0.879 & 0.302 & 0.757 \\
\quad + \texttt{Q2D (ZS)} & 0.687 & 0.892 & 0.662 & 0.894 & 0.350 & 0.781 \\
\quad + \texttt{Q2D (FS)} & 0.690 & 0.886 & \textbf{0.675} & 0.898 & 0.356 & 0.804 \\
\quad + \texttt{Q2D (CoT)} & 0.653 & 0.878 & 0.624 & 0.878 & 0.329 & 0.774 \\
\quad + \texttt{MUGI} & \textbf{0.695} & 0.900 & 0.658 & 0.900 & 0.365 & \textbf{0.822} \\
\midrule
\quad + \texttt{LameR} & 0.637 & 0.857 & 0.653 & \textbf{0.900} & 0.355 & 0.806 \\
\quad + \texttt{CSQE} & 0.690 & \textbf{0.903} & 0.655 & 0.887 & \textbf{0.366} & 0.787 \\
\bottomrule
\end{tabular}}
\vspace{-1em}
\end{table}

\begin{table*}[t]
\centering
\caption{Reproducibility results for LLM-based query reformulation methods using GPT-4.1 with BM25 on BEIR benchmarks. Bold denotes the best score per column. \texttt{MUGI} achieves the highest average on both metrics, followed by corpus-grounded methods (\texttt{LameR}, \texttt{CSQE}) and \texttt{Q2D} variants. \texttt{GenQREnsemble} is particularly effective for recall-oriented tasks.}
\vspace{-1em}
\label{tab:bm25-beir}
\scalebox{0.79}{
\begin{tabular}{l@{\hskip 4pt}c@{\hskip 4pt}c@{\hskip 6pt}c@{\hskip 4pt}c@{\hskip 6pt}c@{\hskip 4pt}c@{\hskip 6pt}c@{\hskip 4pt}c@{\hskip 6pt}c@{\hskip 4pt}c@{\hskip 6pt}c@{\hskip 4pt}c@{\hskip 6pt}c@{\hskip 4pt}c}
\toprule
& \multicolumn{14}{c}{\textbf{BEIR Benchmark}} \\
\cmidrule(lr){2-15}
 & \multicolumn{2}{c}{SciFact} & \multicolumn{2}{c}{ArguAna} & \multicolumn{2}{c}{COVID} & \multicolumn{2}{c}{FiQA} & \multicolumn{2}{c}{DBPedia} & \multicolumn{2}{c}{News} & \multicolumn{2}{c}{Avg.} \\
\cmidrule(lr){2-3} \cmidrule(lr){4-5} \cmidrule(lr){6-7} \cmidrule(lr){8-9} \cmidrule(lr){10-11} \cmidrule(lr){12-13} \cmidrule(lr){14-15}
\textbf{Method} & nDCG@10 & R@100 & nDCG@10 & R@100 & nDCG@10 & R@100 & nDCG@10 & R@100 & nDCG@10 & R@100 & nDCG@10 & R@100 & nDCG@10 & R@100 \\
\midrule
\texttt{Original Query} & 0.679 & 0.925 & 0.300 & 0.932 & 0.595 & 0.109 & 0.236 & 0.539 & 0.318 & 0.468 & 0.395 & 0.447 & 0.421 & 0.626 \\
\midrule
\quad + \texttt{RM3} & 0.646 & 0.915 & 0.286 & 0.955 & 0.593 & 0.117 & 0.192 & 0.497 & 0.308 & 0.459 & 0.426 & 0.519 & 0.412 & 0.646 \\
\quad + \texttt{GenQR} & 0.726 & 0.963 & 0.406 & 0.950 & 0.687 & 0.163 & 0.230 & 0.582 & 0.344 & 0.464 & 0.465 & 0.610 & 0.471 & 0.682 \\
\quad + \texttt{GenQREnsemble} & 0.725 & \textbf{0.967} & 0.407 & \textbf{0.957} & \textbf{0.753} & \textbf{0.184} & 0.239 & 0.580 & 0.360 & 0.476 & 0.486 & \textbf{0.629} & 0.483 & 0.700 \\
\quad + \texttt{Q2K} & 0.709 & 0.940 & 0.406 & 0.938 & 0.715 & 0.177 & \textbf{0.269} & 0.593 & 0.378 & 0.477 & 0.463 & 0.581 & 0.495 & 0.689 \\
\midrule
\quad + \texttt{QA-Expand} & 0.706 & 0.940 & 0.397 & 0.932 & 0.707 & 0.162 & 0.264 & 0.581 & 0.370 & 0.489 & 0.450 & 0.561 & 0.502 & 0.683 \\
\quad + \texttt{Q2D (ZS)} & 0.720 & 0.948 & 0.397 & 0.932 & 0.743 & 0.170 & 0.260 & \textbf{0.600} & 0.406 & 0.505 & 0.498 & 0.586 & 0.525 & 0.701 \\
\quad + \texttt{Q2D (FS)} & 0.712 & 0.949 & 0.401 & 0.941 & 0.708 & 0.164 & 0.268 & 0.599 & 0.401 & 0.508 & 0.480 & 0.584 & 0.521 & 0.704 \\
\quad + \texttt{Q2D (CoT)} & 0.714 & 0.951 & 0.403 & 0.937 & 0.728 & 0.170 & 0.258 & 0.584 & 0.393 & 0.477 & 0.466 & 0.583 & 0.507 & 0.693 \\
\quad + \texttt{MUGI} & \textbf{0.735} & 0.966 & 0.376 & 0.933 & 0.714 & 0.174 & 0.264 & 0.600 & \textbf{0.410} & \textbf{0.531} & \textbf{0.516} & 0.608 & \textbf{0.526} & \textbf{0.715} \\
\midrule
\quad + \texttt{LameR} & 0.725 & 0.949 & \textbf{0.412} & 0.945 & 0.702 & 0.166 & 0.262 & 0.590 & 0.399 & 0.516 & 0.480 & 0.596 & 0.514 & 0.703 \\
\quad + \texttt{CSQE} & 0.721 & 0.949 & 0.398 & 0.945 & 0.699 & 0.164 & 0.247 & 0.584 & 0.390 & 0.514 & 0.479 & 0.591 & 0.516 & 0.703 \\
\bottomrule
\end{tabular}}

\end{table*}

To establish a controlled baseline for cross-method comparison, we evaluate all reformulation methods using GPT-4.1 as the LLM backbone and BM25 as the retriever. This configuration isolates the effect of the reformulation strategy by holding the generation model and retrieval paradigm constant. Alongside the LLM-based methods, we include RM3~\cite{abdul2004umass} as a traditional keyword-level expansion baseline to contextualize LLM-driven gains against classical pseudo-relevance feedback.

\noindent \textbf{TREC DL Benchmark.} Table~\ref{tab:bm25-trec} reports nDCG@10 and Recall@1000 across the three TREC DL collections. All LLM-based methods improve over the original query on DL 2019 and DL 2020, with \texttt{MUGI} achieving the highest nDCG@10 on DL 2019 and \texttt{Q2D (FS)} leading on DL 2020. Document-level expansion methods consistently occupy the top ranks in nDCG@10 across all three collections, confirming that the richer contextual signal provided by pseudo-document generation translates into stronger ranking quality under lexical retrieval. Among the \texttt{Q2D} variants, the chain-of-thought (CoT) configuration consistently underperforms both zero-shot and few-shot, suggesting that intermediate reasoning steps introduce verbosity or tangential content that dilutes the relevance signal rather than enhancing it.

Beyond ranking quality, most LLM-based methods yield substantial improvements in Recall@1000 over the original query across all three collections. On DL 2019, \texttt{CSQE} achieves the highest Recall@1000, while on DL 2020 both \texttt{MUGI} and \texttt{LameR} reach the top recall. On DL-HARD, where the original query retrieves the fewest relevant documents, \texttt{MUGI} improves Recall@1000 by over 14 absolute points. These recall gains indicate that LLM-based reformulations help surface relevant documents that are not retrieved by BM25 with the original query, effectively expanding the candidate pool available to downstream components. In multi-stage retrieval pipelines, this enriched candidate pool can be expected to yield further effectiveness gains when documents are re-ranked by second-stage neural rankers, as a higher proportion of relevant documents in the initial retrieval set directly increases the upper bound on re-ranking performance.

Corpus-grounded methods exhibit particular strength on challenging queries. On DL-HARD, \texttt{CSQE} achieves the best nDCG@10 and \texttt{LameR} the second-best Recall@1000, outperforming several document-level methods that perform well on the easier DL 2019 and DL 2020 collections. This pattern suggests that grounding the expansion in corpus-level signals provides a stabilizing effect when queries are ambiguous or underspecified. In contrast, keyword-level methods show mixed results on DL-HARD: \texttt{GenQREnsemble} degrades nDCG@10 below the original query, indicating that indiscriminate term addition can introduce noise on difficult queries where retrieval precision is critical.

\texttt{RM3} yields only marginal gains on DL 2019 and DL 2020, and notably degrades both nDCG@10 and Recall@1000 on DL-HARD, remaining below the levels achieved by all LLM-based methods. This confirms that classical pseudo-relevance feedback is unreliable when initial retrieval quality is poor, and that LLM-based reformulation offers a clear advantage over feedback-dependent expansion on the TREC DL benchmarks.

\noindent \textbf{BEIR Benchmark.} Table~\ref{tab:bm25-beir} reports nDCG@10 and Recall@100 across the BEIR datasets. \texttt{MUGI} achieves the highest average on both metrics, followed by \texttt{Q2D (ZS)} and \texttt{Q2D (FS)}, reinforcing the dominance of document-level expansion. The advantage is particularly pronounced on datasets with short or underspecified queries such as DBPedia and News, where pseudo-document generation compensates for the limited information in the original query.

Keyword-level methods, while lagging behind document-level approaches in nDCG@10, demonstrate particularly strong recall performance on BEIR. \texttt{GenQREnsemble} achieves the highest Recall@100 on four of the six datasets and attains the best nDCG@10 on COVID, where broad lexical coverage of biomedical terminology appears to be especially beneficial. These recall improvements are noteworthy in the context of multi-stage retrieval architectures, i.e., by bringing a greater number of relevant documents into the top-100 candidate set, keyword-level methods increase the potential for downstream re-rankers to elevate relevant documents into high-rank positions, even when first-stage nDCG@10 gains are modest. Corpus-grounded methods remain competitive on BEIR, with \texttt{LameR} achieving the highest nDCG@10 on ArguAna and both \texttt{CSQE} and \texttt{LameR} ranking among the top methods on average, though their advantage over document-level approaches is less pronounced than on the harder TREC DL queries.

The relative ranking of methods varies substantially across BEIR datasets. FiQA emerges as the most challenging dataset for all methods, with the best-performing approach yielding only a modest gain over the original query. In contrast, ArguAna and COVID exhibit the largest absolute improvements in both nDCG@10 and Recall@100. \texttt{RM3} fails to improve over the original query on average across BEIR, further confirming that classical expansion does not generalize well to heterogeneous domains. This cross-dataset variability highlights the importance of evaluating reformulation methods across diverse domains rather than drawing conclusions from a single benchmark.

% \vspace{-1.0em}
\subsection{RQ2. Cross-Retriever Performance Comparison}

\begin{table*}[t]
\centering
 \caption{Retrieval effectiveness of query reformulation methods generated by GPT-4.1 model across three retrieval paradigms, evaluated on TREC DL and BEIR datasets. Bold values indicate the best score per column within each retriever. LLM-based reformulation yields the largest gains with BM25, offers moderate improvements with BGE (especially on hard queries), but provides limited benefit with SPLADE++, where the original query often remains the strongest.}

\label{tab:results}
\resizebox{\textwidth}{!}{%
\begin{tabular}{l@{\hskip 4pt}c@{\hskip 4pt}c@{\hskip 6pt}c@{\hskip 4pt}c@{\hskip 6pt}c@{\hskip 4pt}c@{\hskip 6pt}c@{\hskip 4pt}c@{\hskip 6pt}c@{\hskip 4pt}c@{\hskip 6pt}c@{\hskip 4pt}c@{\hskip 6pt}c@{\hskip 4pt}c@{\hskip 6pt}c@{\hskip 4pt}c@{\hskip 6pt}c@{\hskip 4pt}c@{\hskip 6pt}c@{\hskip 4pt}c}
\toprule
& \multicolumn{6}{c}{\textbf{TREC DL Benchmark}} & \multicolumn{14}{c}{\textbf{BEIR Benchmark}} \\
\cmidrule(lr){2-7} \cmidrule(lr){8-21}
 & \multicolumn{2}{c}{DL 2019} & \multicolumn{2}{c}{DL 2020} & \multicolumn{2}{c}{DL-HARD} & \multicolumn{2}{c}{SciFact} & \multicolumn{2}{c}{ArguAna} & \multicolumn{2}{c}{COVID} & \multicolumn{2}{c}{FiQA} & \multicolumn{2}{c}{DBPedia} & \multicolumn{2}{c}{News} & \multicolumn{2}{c}{Avg.} \\
\cmidrule(lr){2-3} \cmidrule(lr){4-5} \cmidrule(lr){6-7} \cmidrule(lr){8-9} \cmidrule(lr){10-11} \cmidrule(lr){12-13} \cmidrule(lr){14-15} \cmidrule(lr){16-17} \cmidrule(lr){18-19} \cmidrule(lr){20-21}
\textbf{Method} & nDCG@10 & R@1k & nDCG@10 & R@1k & nDCG@10 & R@1k & nDCG@10 & R@100 & nDCG@10 & R@100 & nDCG@10 & R@100 & nDCG@10 & R@100 & nDCG@10 & R@100 & nDCG@10 & R@100 & nDCG@10 & R@100 \\
\midrule
\multicolumn{21}{l}{\textbf{BM25}} \\
\midrule
\texttt{Original Query} & 0.506 & 0.750 & 0.480 & 0.786 & 0.285 & 0.681 & 0.679 & 0.925 & 0.300 & 0.932 & 0.595 & 0.109 & 0.236 & 0.539 & 0.318 & 0.468 & 0.395 & 0.447 & 0.421 & 0.626 \\
\quad + \texttt{RM3} & 0.522 & 0.814 & 0.490 & 0.824 & 0.251 & 0.716 & 0.646 & 0.915 & 0.286 & 0.955 & 0.593 & 0.117 & 0.192 & 0.497 & 0.308 & 0.459 & 0.426 & 0.519 & 0.412 & 0.646 \\
\quad + \texttt{GenQR} & 0.548 & 0.828 & 0.537 & 0.840 & 0.292 & 0.743 & 0.726 & 0.963 & 0.406 & 0.950 & 0.687 & 0.163 & 0.230 & 0.582 & 0.344 & 0.464 & 0.465 & 0.610 & 0.471 & 0.682 \\
\quad + \texttt{GenQREnsemble} & 0.559 & 0.869 & 0.553 & 0.861 & 0.270 & 0.777 & 0.725 & \textbf{0.967} & 0.407 & \textbf{0.957} & \textbf{0.753} & \textbf{0.184} & 0.239 & 0.580 & 0.360 & 0.476 & 0.486 & \textbf{0.629} & 0.483 & 0.700 \\
\quad + \texttt{QA-Expand} & 0.683 & 0.850 & 0.642 & 0.879 & 0.302 & 0.757 & 0.706 & 0.940 & 0.397 & 0.932 & 0.707 & 0.162 & 0.264 & 0.581 & 0.370 & 0.489 & 0.450 & 0.561 & 0.502 & 0.683 \\
\quad + \texttt{Q2K} & 0.594 & 0.870 & 0.576 & 0.859 & 0.345 & 0.764 & 0.709 & 0.940 & 0.406 & 0.938 & 0.715 & 0.177 & \textbf{0.269} & 0.593 & 0.378 & 0.477 & 0.463 & 0.581 & 0.495 & 0.689 \\
\quad + \texttt{Q2D (ZS)} & 0.687 & 0.892 & 0.662 & 0.894 & 0.350 & 0.781 & 0.720 & 0.948 & 0.397 & 0.932 & 0.743 & 0.170 & 0.260 & \textbf{0.600} & 0.406 & 0.505 & 0.498 & 0.586 & 0.525 & 0.701 \\
\quad + \texttt{Q2D (FS)} & 0.690 & 0.886 & \textbf{0.675} & 0.898 & 0.356 & 0.804 & 0.712 & 0.949 & 0.401 & 0.941 & 0.708 & 0.164 & 0.268 & 0.599 & 0.401 & 0.508 & 0.480 & 0.584 & 0.521 & 0.704 \\
\quad + \texttt{Q2D (CoT)} & 0.653 & 0.878 & 0.624 & 0.878 & 0.329 & 0.774 & 0.714 & 0.951 & 0.403 & 0.937 & 0.728 & 0.170 & 0.258 & 0.584 & 0.393 & 0.477 & 0.466 & 0.583 & 0.507 & 0.693 \\
\quad + \texttt{LameR} & 0.637 & 0.857 & 0.653 & \textbf{0.900} & 0.355 & 0.806 & 0.725 & 0.949 & \textbf{0.412} & 0.945 & 0.702 & 0.166 & 0.262 & 0.590 & 0.399 & 0.516 & 0.480 & 0.596 & 0.514 & 0.703 \\
\quad + \texttt{MUGI} & \textbf{0.695} & 0.900 & 0.658 & 0.900 & 0.365 & \textbf{0.822} & \textbf{0.735} & 0.966 & 0.376 & 0.933 & 0.714 & 0.174 & 0.264 & 0.600 & \textbf{0.410} & \textbf{0.531} & \textbf{0.516} & 0.608 & \textbf{0.526} & \textbf{0.715} \\
\quad + \texttt{CSQE} & 0.690 & \textbf{0.903} & 0.655 & 0.887 & \textbf{0.366} & 0.787 & 0.721 & 0.949 & 0.398 & 0.945 & 0.699 & 0.164 & 0.247 & 0.584 & 0.390 & 0.514 & 0.479 & 0.591 & 0.516 & 0.703 \\
\midrule

\multicolumn{21}{l}{\textbf{SPLADE + +}} \\
\midrule
\texttt{Original Query} & 0.731 & 0.873 & \textbf{0.720} & 0.900 & \textbf{0.385} & \textbf{0.861} & 0.704 & 0.935 & \textbf{0.388} & \textbf{0.984} & \textbf{0.727} & \textbf{0.128} & \textbf{0.347} & 0.631 & \textbf{0.437} & \textbf{0.563} & 0.417 & 0.442 & \textbf{0.539} & 0.702 \\
\quad + \texttt{GenQR} & 0.707 & \textbf{0.933} & 0.626 & 0.914 & 0.380 & 0.849 & \textbf{0.728} & 0.950 & 0.376 & 0.984 & 0.682 & 0.119 & 0.324 & 0.677 & 0.383 & 0.541 & 0.426 & 0.488 & 0.514 & \textbf{0.717} \\
\quad + \texttt{GenQREnsemble} & 0.686 & 0.902 & 0.586 & 0.914 & 0.305 & 0.821 & 0.718 & 0.943 & 0.381 & 0.981 & 0.673 & 0.120 & 0.301 & 0.654 & 0.364 & 0.536 & 0.444 & \textbf{0.505} & 0.495 & 0.709 \\
\quad + \texttt{QA-Expand} & \textbf{0.734} & 0.917 & 0.674 & 0.926 & 0.355 & 0.803 & 0.696 & 0.949 & 0.382 & 0.980 & 0.694 & 0.115 & 0.340 & 0.682 & 0.387 & 0.529 & 0.427 & 0.457 & 0.521 & 0.707 \\
\quad + \texttt{Q2K} & 0.681 & 0.930 & 0.652 & 0.925 & 0.352 & 0.838 & 0.719 & 0.939 & 0.382 & 0.981 & 0.687 & 0.122 & 0.328 & 0.667 & 0.394 & 0.548 & 0.421 & 0.499 & 0.513 & 0.717 \\
\quad + \texttt{Q2D (ZS)} & 0.700 & 0.914 & 0.688 & 0.937 & 0.338 & 0.839 & 0.704 & 0.955 & 0.382 & 0.981 & 0.634 & 0.109 & 0.330 & 0.677 & 0.395 & 0.521 & 0.452 & 0.479 & 0.513 & 0.712 \\
\quad + \texttt{Q2D (FS)} & 0.693 & 0.907 & 0.675 & 0.939 & 0.377 & 0.840 & 0.709 & 0.957 & 0.383 & 0.981 & 0.659 & 0.110 & 0.345 & \textbf{0.689} & 0.391 & 0.519 & 0.430 & 0.501 & 0.518 & 0.716 \\
\quad + \texttt{Q2D (CoT)} & 0.688 & 0.915 & 0.653 & 0.909 & 0.331 & 0.846 & 0.712 & 0.946 & 0.382 & 0.980 & 0.686 & 0.106 & 0.315 & 0.651 & 0.393 & 0.532 & 0.416 & 0.474 & 0.508 & 0.707 \\
\quad + \texttt{LameR} & 0.684 & 0.906 & 0.639 & 0.938 & 0.367 & 0.825 & 0.718 & 0.958 & 0.384 & 0.983 & 0.631 & 0.108 & 0.329 & 0.672 & 0.356 & 0.490 & \textbf{0.452} & 0.477 & 0.507 & 0.706 \\
\quad + \texttt{MUGI} & 0.686 & 0.909 & 0.651 & 0.920 & 0.362 & 0.811 & 0.706 & \textbf{0.960} & 0.370 & 0.978 & 0.646 & 0.112 & 0.335 & 0.680 & 0.384 & 0.514 & 0.442 & 0.500 & 0.509 & 0.709 \\
\quad + \texttt{CSQE} & 0.694 & 0.919 & 0.680 & \textbf{0.940} & 0.369 & 0.834 & 0.707 & 0.959 & 0.380 & 0.983 & 0.681 & 0.112 & 0.329 & 0.675 & 0.396 & 0.523 & 0.450 & 0.502 & 0.521 & 0.716 \\
\midrule

\multicolumn{21}{l}{\textbf{BGE-base-en-v1.5}} \\
\midrule
\texttt{Original Query} & 0.702 & 0.843 & 0.677 & 0.855 & 0.178 & 0.231 & 0.741 & 0.967 & \textbf{0.636} & \textbf{0.992} & 0.781 & 0.141 & 0.406 & 0.742 & 0.407 & 0.530 & 0.442 & 0.499 & 0.528 & 0.625 \\
\quad + \texttt{GenQR} & 0.702 & 0.865 & 0.690 & 0.852 & 0.387 & 0.840 & 0.748 & 0.970 & 0.626 & 0.989 & 0.778 & \textbf{0.147} & 0.392 & 0.733 & 0.355 & 0.469 & 0.464 & 0.509 & 0.572 & 0.708 \\
\quad + \texttt{GenQREnsemble} & 0.703 & 0.887 & 0.683 & 0.870 & 0.357 & 0.863 & 0.759 & 0.970 & 0.619 & 0.990 & 0.800 & 0.144 & 0.403 & 0.746 & 0.376 & 0.496 & 0.475 & \textbf{0.525} & 0.575 & 0.721 \\
\quad + \texttt{QA-Expand} & 0.737 & 0.894 & 0.707 & 0.875 & 0.374 & 0.854 & 0.737 & 0.960 & 0.623 & 0.990 & 0.795 & 0.142 & 0.416 & 0.745 & 0.401 & 0.509 & 0.470 & 0.485 & 0.584 & 0.717 \\
\quad + \texttt{Q2K} & 0.697 & 0.870 & 0.642 & 0.818 & 0.378 & 0.831 & 0.742 & 0.963 & 0.619 & 0.990 & 0.774 & 0.140 & 0.392 & 0.741 & 0.325 & 0.427 & 0.445 & 0.485 & 0.557 & 0.696 \\
\quad + \texttt{Q2D (ZS)} & 0.728 & 0.899 & \textbf{0.739} & \textbf{0.906} & 0.379 & 0.859 & \textbf{0.761} & 0.963 & 0.619 & 0.990 & \textbf{0.806} & 0.145 & 0.415 & 0.749 & 0.431 & 0.522 & 0.476 & 0.511 & 0.595 & 0.727 \\
\quad + \texttt{Q2D (FS)} & 0.727 & 0.889 & 0.714 & 0.895 & 0.407 & \textbf{0.873} & 0.752 & 0.967 & 0.618 & 0.989 & 0.804 & 0.141 & 0.420 & 0.754 & 0.430 & \textbf{0.530} & 0.471 & 0.516 & 0.594 & \textbf{0.728} \\
\quad + \texttt{Q2D (CoT)} & 0.713 & 0.888 & 0.672 & 0.876 & 0.376 & 0.851 & 0.758 & 0.963 & 0.619 & 0.989 & 0.798 & 0.138 & 0.401 & 0.748 & 0.368 & 0.456 & 0.433 & 0.476 & 0.571 & 0.709 \\
\quad + \texttt{LameR} & 0.703 & 0.889 & 0.715 & 0.903 & 0.412 & 0.856 & 0.757 & 0.973 & 0.620 & 0.989 & 0.780 & 0.137 & 0.408 & 0.741 & 0.402 & 0.500 & 0.437 & 0.459 & 0.582 & 0.716 \\
\quad + \texttt{MUGI} & 0.735 & 0.887 & 0.720 & 0.895 & 0.404 & 0.842 & 0.757 & \textbf{0.977} & 0.616 & 0.990 & 0.802 & 0.143 & \textbf{0.429} & \textbf{0.758} & \textbf{0.440} & 0.529 & \textbf{0.490} & 0.521 & \textbf{0.599} & 0.727 \\
\quad + \texttt{CSQE} & \textbf{0.755} & \textbf{0.901} & 0.714 & 0.897 & \textbf{0.414} & 0.864 & 0.755 & 0.963 & 0.622 & 0.992 & 0.788 & 0.143 & 0.407 & 0.738 & 0.424 & 0.523 & 0.463 & 0.507 & 0.594 & 0.725 \\
\bottomrule
\end{tabular}}

\end{table*}

To assess whether reformulation gains extend beyond lexical retrieval, we evaluate all methods across BM25, SPLADE, and BGE while keeping the reformulation model and decoding configuration fixed. This design isolates the interaction between reformulation strategy and retrieval paradigm, enabling direct comparison of how expanded queries behave across different representation spaces.

\noindent \textbf{TREC DL Benchmark.} Table~\ref{tab:results} reports nDCG@10 and Recall@1000 across the three retrieval paradigms. Under BM25, reformulation consistently improves both ranking quality and recall, as discussed in RQ1. However, this pattern does not uniformly extend to learned sparse and dense retrievers. For SPLADE, gains are smaller and less consistent. Several document-level methods that substantially improve BM25 exhibit only marginal improvements under SPLADE, and in some cases introduce slight degradations in nDCG@10. This suggests partial redundancy in that SPLADE already performs neural term expansion during encoding, reducing the marginal benefit of explicit LLM-generated lexical enrichment. Keyword-level methods show even more limited gains under SPLADE, indicating that additional term injection may overlap with SPLADE’s learned expansion signals rather than introducing complementary information.

Under dense retrieval with BGE, the divergence is more pronounced. While a subset of document-level methods maintains modest improvements on DL 2019 and DL 2020, others fail to outperform the original query. In some cases, improvements in recall are limited or marginal relative to the baseline. Because dense retrieval operates in a continuous embedding space, surface-level lexical augmentation does not necessarily translate into proportional embedding shifts. Expansions that improve term matching for BM25 may therefore alter the semantic embedding in directions that are not aligned with relevant documents in vector space.

Notably, corpus-grounded methods demonstrate relatively stronger transfer to SPLADE and BGE on DL-HARD compared to purely generative document-level approaches. Conditioning on initially retrieved evidence appears to constrain the expansion to distributionally relevant terms, mitigating embedding drift and preserving alignment with the retriever’s representation space. In contrast, unconstrained pseudo-document synthesis occasionally introduces topical broadening that benefits lexical recall but perturbs dense similarity.

\noindent \textbf{BEIR Benchmark.} Table~\ref{tab:results} also reports average nDCG@10 and Recall@100 across the six BEIR datasets. The cross-retriever pattern observed on TREC DL generalizes to heterogeneous domains. Under BM25, document-level methods achieve the strongest average gains. Under SPLADE, improvements are reduced and cluster more tightly across methods. Under BGE, performance differences narrow further, with several methods yielding negligible or negative changes relative to the original query.

Dataset-level variability remains substantial. On term-sensitive collections such as SciFact and COVID, lexical retrieval benefits most from expansion, whereas dense retrieval exhibits more modest sensitivity. On semantically oriented datasets such as ArguAna, dense retrieval already captures high-level meaning effectively, and additional generative expansion provides limited benefit. In some cases, expanded queries introduce verbosity that dilutes embedding specificity, slightly reducing nDCG@10.

Across both benchmark families, the key observation is that reformulation effectiveness is strongly conditioned on the retriever’s representational assumptions. Methods that appear consistently effective under BM25 do not uniformly transfer to learned sparse or dense retrieval. The magnitude of improvement systematically decreases as the retriever moves from exact term matching to neural encoding with implicit expansion.

These findings indicate that LLM-based query reformulation should not be evaluated exclusively under lexical retrieval. Improvements observed in BM25 settings cannot be assumed to generalize to neural retrievers. Instead, the interaction between expansion strategy and retrieval representation space emerges as a primary determinant of effectiveness. This interaction explains part of the variability in prior literature, where reformulation gains reported under lexical baselines may not reflect cross-paradigm robustness. At the same time, a practically significant observation emerges: BM25, when coupled with effective LLM-based expansion, frequently approaches or matches the retrieval effectiveness of dense retrievers operating on unexpanded queries. This performance is achieved without the substantial overhead required for constructing vector-based index, training neural retrievers, or fine-tuning encoders. These findings suggest that LLM-based query reformulation over a standard lexical index may serve as a competitive and cost-effective alternative to dense retrieval methods. All configurations enabling this comparison are released as runnable pipelines in QueryGym.

\begin{figure*}[t]
\centering

\begin{minipage}{0.68\textwidth}
    \centering
    \includegraphics[width=\linewidth]
    {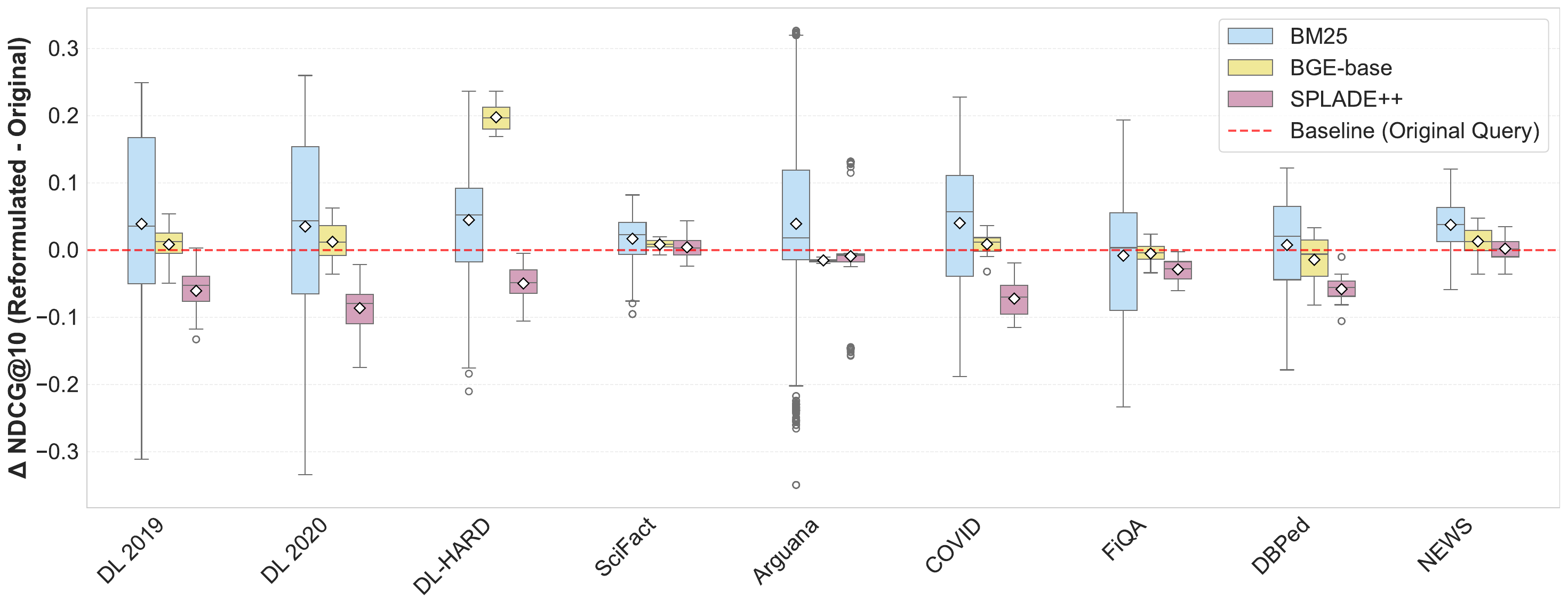}
    \caption*{(a)}
\end{minipage}
\hfill
\begin{minipage}{0.30\textwidth}
    \centering
    \includegraphics[width=\linewidth]
    {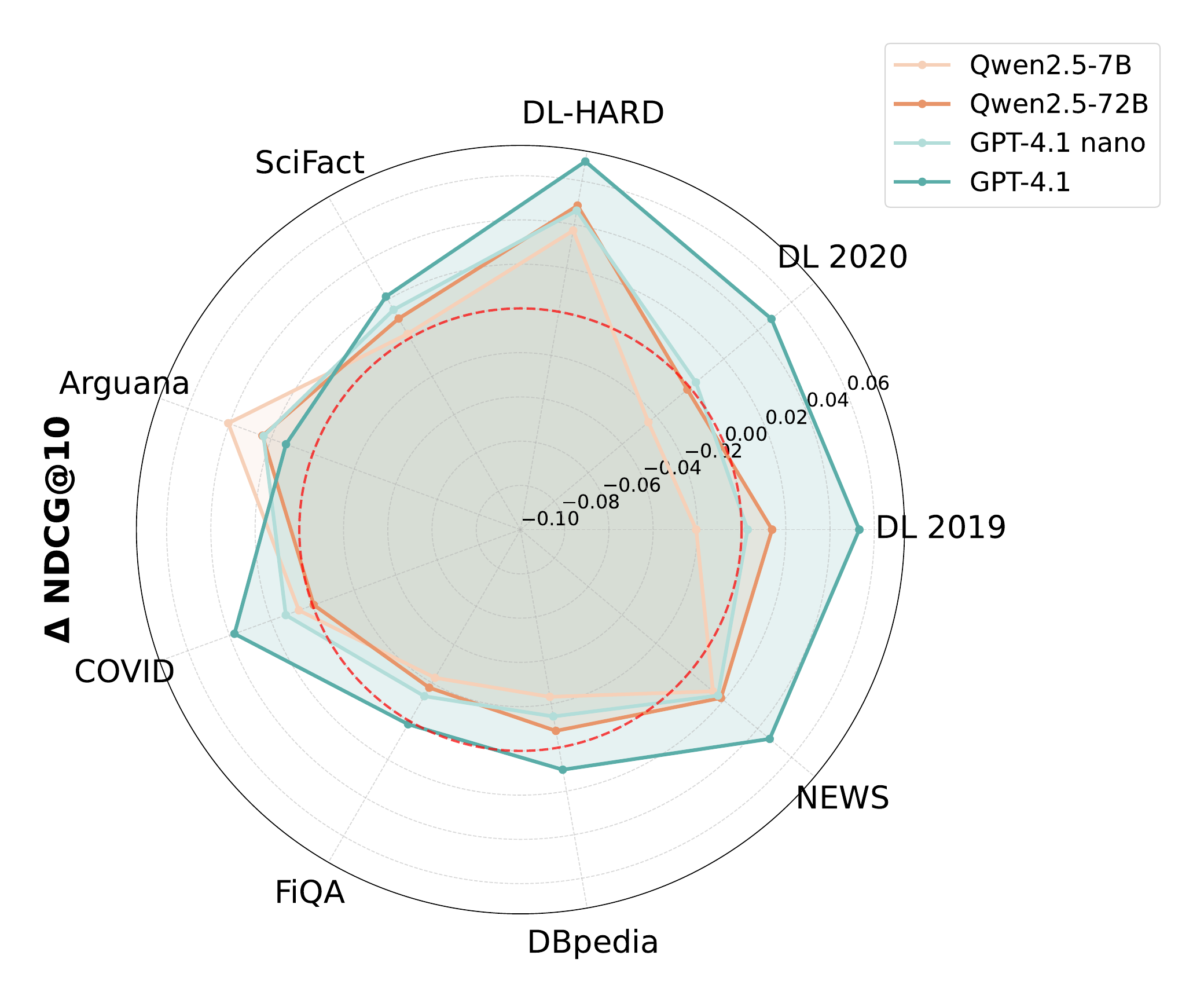}
    \caption*{(b)}
\end{minipage}

\vspace{-0.5em}
\caption{Domain–level effectiveness variation of LLM–based query reformulation. 
(a) Distribution of per–query $\Delta$ nDCG@10 relative to the original query across three retrieval paradigms, where all reformulations are generated using GPT-4.1. Each box summarizes the gain or loss induced by reformulation within a dataset, and the dashed horizontal line denotes the no–change baseline, highlighting both positive improvements and negative regressions. 
(b) Aggregate $\Delta$ nDCG@10 across datasets for different LLM backbones and parameter scales, illustrating cross–domain performance trends and model scale sensitivity.}

\label{fig:RQ3_combined}
\end{figure*}

\subsection{RQ3. Domain Robustness and Dataset Sensitivity}

To assess how consistently reformulation methods generalize across heterogeneous domains, we analyze the distribution of nDCG@10 gains ($\Delta$ nDCG@10 = reformulated $-$ original) across all nine evaluation datasets. Figure~\ref{fig:RQ3_combined}(a) presents box plots of these gains grouped by retrieval paradigm, aggregating over all reformulation methods, while Figure~\ref{fig:RQ3_combined}(b) provides a radar chart of the same metric grouped by LLM backbone.

\noindent \textbf{Retriever-level domain sensitivity.} Figure~\ref{fig:RQ3_combined}(a) reveals a clear hierarchy in how retrieval paradigms respond to reformulation across domains. BM25 exhibits the largest and most consistently positive gains, with median $\Delta$ nDCG@10 above the baseline on all nine datasets. However, the variance of these gains differs substantially by domain. DL 2019, DL 2020, and DL-HARD show wide interquartile ranges, reflecting that some reformulation methods yield large improvements while others provide minimal benefit. Arguana stands out as the most volatile dataset under BM25, with individual methods spanning from gains exceeding +0.30 to degradations approaching $-$0.35, indicating that this argumentative retrieval task is highly sensitive to the nature of the expansion content.

BGE displays a qualitatively different profile. On most datasets, the median $\Delta$ nDCG@10 lies near or slightly below zero, suggesting limited consistent benefit for dense retrieval. The striking exception is DL-HARD, where BGE achieves the largest positive median gain of any retriever-dataset combination, indicating that generative expansion can meaningfully enrich semantic representations even in embedding space when queries are underspecified. On COVID and DBPedia, however, BGE medians fall below the baseline, suggesting that lexical augmentation can shift the query embedding away from the relevant region when the original query already provides adequate semantic coverage.

SPLADE exhibits the most compressed distributions across all datasets, with medians consistently near zero and narrow interquartile ranges. This confirms that SPLADE's built-in neural term expansion absorbs much of the benefit that explicit reformulation provides to BM25, leaving minimal room for further query-side enrichment. On DL 2019, DL 2020, and DL-HARD, SPLADE medians fall slightly below zero, indicating that additional expansion can introduce degradation when the retriever already performs implicit expansion.

\noindent \textbf{LLM-level domain sensitivity.} Figure~\ref{fig:RQ3_combined}(b) reveals systematic differences in how LLM backbones distribute reformulation gains across datasets. The GPT-4.1 family produces consistently larger radar polygons, indicating positive $\Delta$ nDCG@10 across the majority of datasets. GPT-4.1 achieves the strongest improvements on DL-HARD, DL 2019, DL 2020, and News, while maintaining positive gains on all remaining collections. GPT-4.1-nano traces a remarkably similar profile, trailing only marginally on individual datasets. This narrow within-family gap suggests that for the GPT-4.1 architecture, even the compact variant generates expansion content of sufficient quality to consistently benefit retrieval.

The Qwen family shows a smaller overall footprint and higher scale sensitivity. Qwen2.5-72B yields modest positive gains on most TREC DL collections but approaches or falls below the baseline on COVID and FiQA, while Qwen2.5-7B exhibits the weakest profile, with $\Delta$ nDCG@10 dropping below zero on COVID and remaining near zero elsewhere. The larger performance gap between Qwen2.5-7B and Qwen2.5-72B, relative to the GPT-4.1 pair, suggests that reformulation quality is family-dependent rather than following a universal scaling trend.

Across all four LLMs, DL-HARD consistently receives the largest positive gains, confirming that reformulation provides the greatest benefit on challenging, underspecified queries regardless of the backbone model. Conversely, FiQA remains the most resistant dataset to reformulation, with all four LLMs producing only marginal improvements. COVID and ArguAna exhibit the greatest cross-model variability: stronger models generate effective expansions on these collections, while weaker models introduce noise that degrades performance below the original query baseline.

These findings indicate that domain robustness is not an intrinsic property of any single reformulation method or model but rather emerges from the interaction between the expansion strategy, the retrieval paradigm, and the LLM backbone. Evaluations conducted on a single dataset, retriever, or model risk overstating the generalizability of reported gains.

\begin{figure}[t]
    \centering
    \includegraphics[width=\columnwidth]{figures/rank_consistency.png}
    \caption{Rank coefficient of variation (RankCV) for all method--LLM combinations across datasets. Lower RankCV indicates methods whose \emph{relative} ranking is stable across datasets (generalists), while higher RankCV indicates methods whose ranking depends strongly on the dataset (specialists). Ranks are computed within each LLM configuration on a 1--$|M|$ scale.}
    \vspace{-1.5em}
    \label{fig:rank_consistency}
\end{figure}

\begin{figure}[t]
    \centering
    \includegraphics[width=\columnwidth]{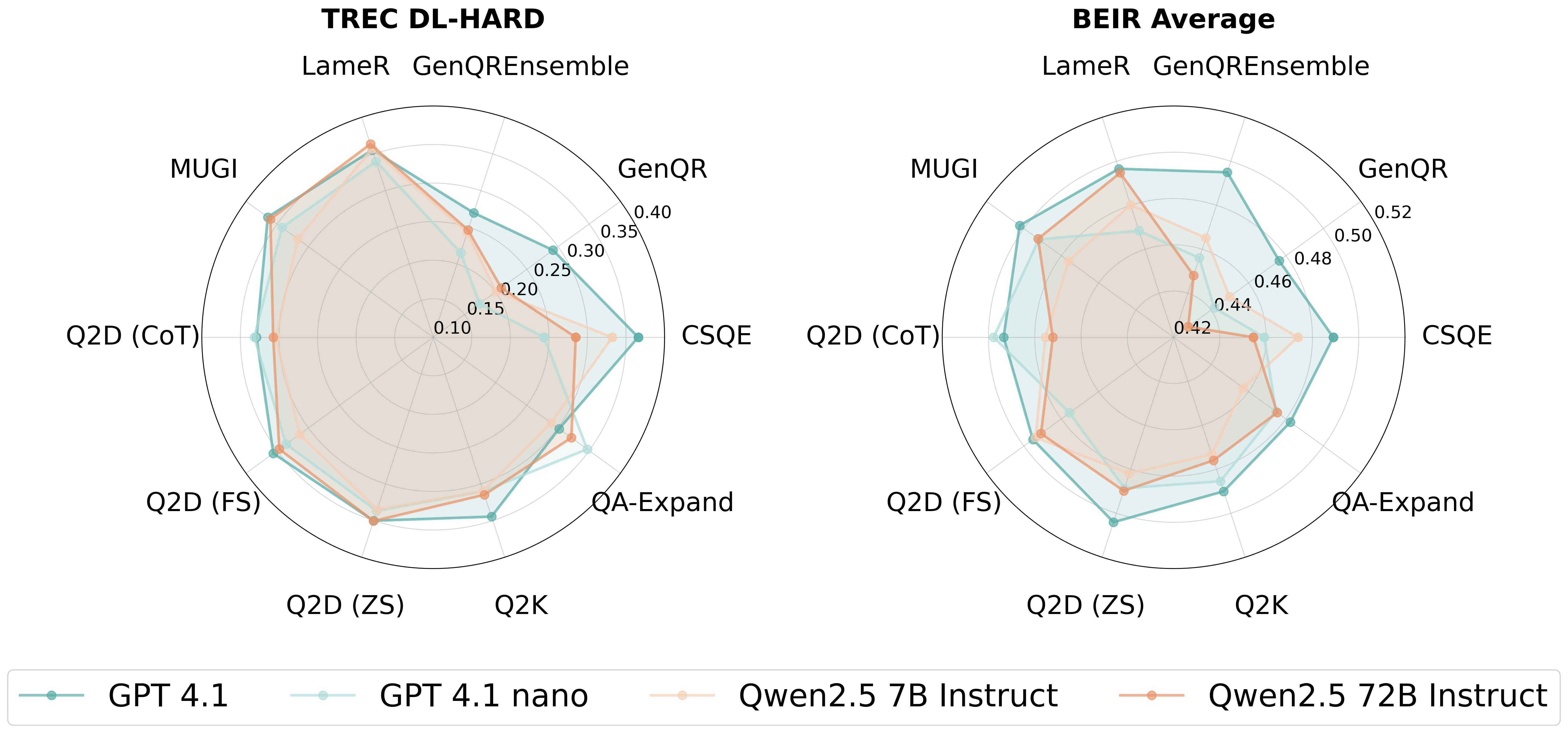}
    \caption{Radar charts comparing method performance across four LLM backbones on TREC DL-HARD (left) and BEIR Average (right). Each axis represents one method; radial distance indicates nDCG@10 performance. Crossing lines indicate ranking instability. Notably, GPT 4.1 and GPT 4.1 nano show low rank correlation ($\rho = 0.35$--$0.52$, ns), while Qwen 72B and 7B maintain stable rankings ($\rho = 0.77$--$0.86$, $p < 0.01$). LLM $\times$ method interaction explains 16--22\% of total variance, comparable to LLM main effects (8--23\%).}
    \vspace{-1.0em}
    \label{fig:llm_interaction}
\end{figure}

\vspace{-0.5em}
\subsection{RQ4. Impact of LLM Backbone and Model Scale}
\label{sec:rq4}
\subsubsection{Ranking stability across LLM backbones and scales}
We next examine whether the \emph{relative} ordering of reformulation strategies is stable when we vary the LLM backbone and parameter scale. While prior work often reports gains for a method under a single LLM configuration, reproducibility across backbones and scales is less clear: \textit{a strategy that performs well with one model may not preserve its advantage when the underlying generator changes.} To quantify this stability independently of absolute effectiveness and dataset difficulty, we analyze \emph{rank consistency} across datasets within each LLM configuration.

\noindent \textbf{Rank consistency metric.}
For each dataset $d \in \mathcal{D}$ and LLM configuration $\ell$ (backbone $\times$ scale), we rank the $|M|$ reformulation methods by nDCG@10 from 1 (best) to $|M|$ (worst). Ranks are computed within each $\ell$ to avoid conflating changes in absolute score levels across LLMs with changes in relative method ordering. For each method--LLM pair $(m,\ell)$, we then compute the coefficient of variation (CV) of its ranks across datasets:
\begin{equation}
\text{RankCV}(m,\ell) = \frac{\sigma(\{r_{d,\ell}(m)\}_{d \in \mathcal{D}})}{\mu(\{r_{d,\ell}(m)\}_{d \in \mathcal{D}})} \times 100\%.
\end{equation}

Lower RankCV indicates a \emph{generalist} strategy whose relative standing is stable across domains, whereas higher RankCV indicates a \emph{specialist} that alternates between top and bottom ranks depending on the dataset.

Figure~\ref{fig:rank_consistency} summarizes RankCV for all method--LLM combinations. The distribution is broad (median 40.4\%), indicating that cross-domain stability is not the norm: many strategies change substantially in relative ranking as the evaluation domain varies, even under a fixed LLM.

First, we observe that the most \emph{consistent} combinations tend to occupy mid-tier ranks rather than the top of the leaderboard. For example, \texttt{QA-Expand} (Qwen2.5 7B), \texttt{Q2D (CoT)} (Qwen2.5 7B), and \texttt{GenQREnsemble} (GPT-4.1 nano) exhibit the lowest RankCV values in Figure~\ref{fig:rank_consistency}, but their ranks remain largely in the middle-to-lower range across datasets. This pattern suggests that some strategies offer predictable behavior across domains, but do not necessarily deliver best-in-class effectiveness on any particular collection.

Second, several of the strongest-performing strategies in our earlier analyses exhibit pronounced \emph{domain specialization}. In particular, \texttt{MUGI} shows extreme variability in relative rank across datasets across all evaluated LLMs (i.e., it can be among the top-ranked methods on some datasets and near the bottom on others). Importantly, this specialization pattern persists across backbones and scales, suggesting that it is primarily method-intrinsic rather than an artifact of a specific LLM configuration.

\noindent \textbf{Backbone and scale effects.}
Rank consistency differs across LLM configurations, but the dominant driver is the reformulation \emph{family} rather than the generator choice. For example, \texttt{GenQR}/\texttt{GenQREnsemble} variants are consistently among the more stable strategies, whereas \texttt{MUGI} variants are consistently among the least stable. Scale effects are method dependent: some strategies become more stable with larger models while others become less stable, reinforcing our broader finding that improvements from increasing LLM capacity are not uniform across reformulation approaches or domains.

\begin{table}[t]
\centering
\caption{Spearman rank correlations between LLM pairs. \\ ** $p < 0.01$, * $p < 0.05$, ns = not significant.}

\label{tab:rank_correlations}
\small
\begin{tabular}{lcc}
\toprule
\textbf{LLM Pair} & \textbf{DL-HARD} & \textbf{BEIR Avg} \\
\midrule
GPT-4.1 $\leftrightarrow$ GPT-4.1 nano & 0.345 (ns) & 0.515 (ns) \\
GPT-4.1 $\leftrightarrow$ Qwen2.5 72B & 0.758** & 0.842** \\
GPT-4.1 $\leftrightarrow$ Qwen2.5 7B & 0.806** & 0.830** \\
GPT-4.1 nano $\leftrightarrow$ Qwen2.5 72B & 0.661* & 0.673* \\
GPT-4.1 nano $\leftrightarrow$ Qwen2.5 7B & 0.467 (ns) & 0.539 (ns) \\
Qwen2.5 72B $\leftrightarrow$ Qwen2.5 7B & 0.855** & 0.770** \\
\bottomrule
\end{tabular}

\end{table}

\begin{table}[t]
\centering
\caption{Variance partitioning of nDCG@10 across reformulation methods and LLMs. Numbers show the percentage of score variation explained by LLM effects, method effects, and LLM × method interactions (plus residual).}

\label{tab:variance_partition}
\small
\begin{tabular}{lcc}
\toprule
\textbf{Variance Source} & \textbf{DL-HARD} & \textbf{BEIR Avg} \\
\midrule
LLM main effect & 8.4\% & 23.3\% \\
Method main effect & 75.7\% & 54.5\% \\
Interaction + residual & 16.0\% & 22.2\% \\
\midrule
Additive model (LLM+Method) & 84.0\% & 77.8\% \\
\bottomrule
\end{tabular}
\vspace{-1em}
\end{table}

\subsubsection{Rank Agreement Between LLM Pairs}
We test a common implicit assumption in LLM-based reformulation studies: \emph{method rankings generalize across LLM backbones and scales}. If this assumption held, benchmarking a reformulation strategy with one generator would largely predict its relative standing with other LLMs. Conversely, if rankings are unstable, comparative claims (e.g., ``method $A$ outperforms method $B$'') must be scoped to a specific LLM configuration, limiting portability and reproducibility.

\noindent \textbf{Rank correlation.}
For DL-HARD and BEIR (averaged across datasets), we rank methods within each LLM configuration by nDCG@10 and compute pairwise Spearman rank correlations between LLMs. Higher $\rho$ indicates that the ordering of methods is preserved when changing the generator; lower $\rho$ indicates LLM-dependent rank reversals. Table~\ref{tab:rank_correlations} summarizes the correlations.

\noindent \textbf{Variance partitioning.}
We further quantify LLM sensitivity by decomposing observed performance variance into (i) an LLM main effect, (ii) a method main effect, and (iii) an interaction term capturing LLM $\times$ method dependence. A large interaction component implies that method effectiveness cannot be explained by additive ``better method'' and ``better LLM'' effects alone.

\noindent \textbf{Results.}
Table~\ref{tab:rank_correlations} indicates that within-family stability differs substantially. In particular, the rank correlation between GPT-4.1 and GPT-4.1 nano is low and not statistically significant on either benchmark, whereas Qwen2.5 72B and Qwen2.5 7B preserve method rankings much more strongly. Cross-family correlations are generally moderate to strong when comparing full-scale models, but weaker and less consistent when GPT-4.1 nano is involved. This may be because GPT-4.1 nano is relatively small and less capable, leading to noisier reformulations that amplify ranking instability. Figure~\ref{fig:llm_interaction} provides a qualitative view of these effects: crossing lines correspond to rank reversals across LLMs.

Finally, Table~\ref{tab:variance_partition} shows that method choice explains the majority of variance. The \emph{LLM main effect} captures score differences attributable to changing the generator while averaging over reformulation methods; the \emph{method main effect} captures differences attributable to changing the reformulation strategy while averaging over LLMs. The \emph{interaction} component captures non-additive LLM $\times$ method dependence (i.e., cases where a method’s advantage changes with the LLM, leading to rank reversals), plus residual variability not explained by the two main effects.
Under this decomposition, method choice explains most variation on both benchmarks (54.5--75.7\%), but the interaction term remains substantial (16.0--22.2\%), indicating that relative conclusions about methods are not fully portable across LLM configurations even when overall LLM quality differs only modestly. Taken together, these results imply that LLM selection is part of the experimental condition for reformulation benchmarking, and that reproducibility claims based on a single generator (especially a single scale) should be interpreted with caution.

\noindent \textbf{Implications for reproducibility.}
Together with our cross-domain stability analysis (RankCV), these findings highlight a limitation of reporting only mean effectiveness across datasets: the same average score can arise from a reliable generalist (moderate ranks everywhere) or an unstable specialist (very high on some datasets and very low on others). Moreover, non-trivial LLM $\times$ method interactions imply that comparative claims (e.g., ``method $A$ outperforms method $B$'') may not transfer across generators or scales. For reproducible conclusions and actionable guidance, evaluations should therefore report both average effectiveness and stability (e.g., RankCV) across datasets, and should validate conclusions across multiple LLM configurations whenever possible.

\section{Concluding Remarks}
This study revisited LLM-based query reformulation under a strictly controlled experimental framework designed to isolate the contribution of the reformulation strategy from confounding factors such as decoding configuration, model backbone, retrieval method, and dataset choice. By evaluating ten representative methods across multiple LLM families, parameter scales, retrieval architectures, and benchmark collections, we provide a systematic assessment of which previously reported gains are stable and which are configuration-dependent.

Our findings indicate that improvements observed under lexical retrieval do not consistently transfer to learned sparse or dense retrievers, and that the relationship between model scale and downstream effectiveness is neither uniform nor monotonic. Reformulation gains vary substantially across domains and query difficulty, and comparative method rankings are not preserved across retrieval methods. These results suggest that conclusions drawn from single-retriever or single-dataset evaluations should be interpreted cautiously. By releasing all prompts, configurations, and evaluation artifacts, we aim to provide a transparent reference framework for future studies and to support more standardized and comparable evaluation practices. The QueryGym toolkit and leaderboard provide a continuing venue where new reformulation methods, LLM backbones, and retrieval paradigms can be evaluated against the configurations established in this study.

\balance
\bibliographystyle{ACM-Reference-Format}
\bibliography{references}

\end{document}